\documentclass[]{article}
\usepackage{amsmath,amssymb,amsthm}
\usepackage{bm}
\usepackage{dsfont}
\usepackage[T1]{fontenc}
\usepackage[utf8]{inputenc}
\usepackage[english]{babel}
\usepackage{indentfirst}
\usepackage{xcolor}
\usepackage[colorlinks = true, linkcolor = black]{hyperref}
\usepackage{geometry}
\usepackage{graphicx}
\usepackage{rotating}
\usepackage{subcaption}
\usepackage{verbatim}
\usepackage[titletoc]{appendix}
\usepackage{booktabs}
\usepackage{tikz}
\usepackage[ruled,vlined,resetcount,linesnumbered]{
algorithm2e}
\usepackage{textcomp}
\usepackage{subfiles}
\usepackage{stmaryrd}
\usepackage{enumitem}

\newtheorem{theorem}{Theorem}
\newtheorem{definition}[theorem]{Definition}
\newtheorem{proposition}[theorem]{Proposition}

\newtheorem{remark}[theorem]{Remark}

\newcommand{\etal}{\emph{et al.}}
\newcommand{\tm}{\textsuperscript{TM} }

\tikzstyle{state}=[shape=circle,draw=blue!50,fill=blue!20]
\tikzstyle{observation}=[shape=rectangle,draw=orange!50,fill=orange!20]
\tikzstyle{lightedge}=[<-,dotted]
\tikzstyle{mainstate}=[state,thick]
\tikzstyle{mainedge}=[<-,thick]

\begin{document}

\title{Real-time public transport service-level monitoring using passive WiFi: a spectral clustering approach for train timetable estimation}
\author{Baoyang Song\\Ecole Polytechnique\\France\\ baoyang.song@polytechnique.edu \and Laura Wynter\\IBM Research\\ Singapore\\lwynter@sg.ibm.com}
\date{}
  \maketitle

    \begin{abstract}
       A new area in which passive WiFi analytics have promise for  delivering  value is the real-time  monitoring  of public transport systems. 
 One  example is determining the true (as opposed to the published) timetable of a public transport system in real-time. In most cases, there are no other publicly-available sources for this information. Yet, it is indispensable for the real-time monitoring of public transport service levels. Furthermore, this information, if accurate and temporally fine-grained, can be used for very low-latency incident detection.
In this work, we propose using spectral clustering based on trajectories derived from  passive WiFi traces of users of a public transport system to infer the true timetable and  two key performance indicators of the transport service, namely public transport vehicle headway and in-station dwell time. By detecting anomalous dwell times or headways, we demonstrate that a fast and accurate real-time incident-detection procedure can be obtained. The method we introduce makes use of the advantages of the high-frequency WiFi data, which provides very low-latency, universally-accessible information, while minimizing the impact of the noise in the data. 
    \end{abstract}

  \subsection*{Keywords} online  monitoring, incident detection, rail, machine learning

\section{Introduction}
\label{sec:intro}

WiFi access points (AP) are  omnipresent 
and most  mobile stations (cell phones, laptops, e-readers, etc.) today are 
equipped with WiFi functionality. 
In order to discover and automatically connect to known WiFi networks, 
mobile stations scan periodically the WiFi bands by broadcasting on all 
available channels  so-called 
\emph{probe 
requests}\cite{Musa:2012:TUS:2426656.2426685,cunche:hal-00874078,
Barbera:2013:SCU:2504730.2504742,paper-51}. 
Furthermore, even after
associating successfully with an AP, 
mobile stations continue sending probe requests \cite{Musa:2012:TUS:2426656.2426685}. This happens especially 
when the connection is unstable or when user is roaming between different APs. 
Probe requests, specified by the IEEE 802.11 
protocol\cite{dot11}, carry valuable information such as the MAC address of the
sending device, the signal strength, etc.
It is interesting to note that a mobile station's probe requests are universally accessible. While administrators of 
APs can simply query the system log, anyone can access probe requests sent 
by mobile stations using a  WiFi sniffer 
such as \texttt{tcpdump} or \texttt{Wireshark}.
Probe requests are not bound to any specific AP as they occur before the association with APs. Thus, even if no APs are present, probe requests are 
 sent by individuals' devices and can still be observed.
 In addition,  accessing probe requests is device-free and non-intrusive. No 
 hardware modules are needed on the system side and no software need be installed 
 on the individual's mobile station.

Because of these clear advantages, real-time WiFi analytics have  been gaining 
interest among researchers in recent years. 
Handte \etal\cite{paper-51} designed  a system to estimate the 
number of passengers in public transport vehicles. Musa 
\etal\cite{Musa:2012:TUS:2426656.2426685} described how to exploit probe 
requests to infer trajectory of vehicles by using an Hidden Markov Model (HMM). 
Bonné \etal\cite{6583443} built a system on top of a Raspberry Pi\tm to track 
users' location at a mass event using probe requests, association requests and 
reassociation requests.
Wang \etal\cite{Wang:2014:THQ:2594368.2594382} studied queue waiting time 
measurement 
using a single-point WiFi monitoring approach and designed queue measurement 
techniques adaptive to different period of time based on 
Bayesian networks. 
Manweiler \etal\cite{6567123} designed a dwell time prediction framework in 
retail store environments using various sensors from smartphones including WiFi 
signals strength and data transmission rate. This work was extended to predict 
the length of stay of patrons in e.g. a retail environment using a passive WiFi 
sensing system and an SVM with  online learning  in \cite{viet_paper}.

 In an offline setting, Rose \etal\cite{Rose:2010:MUW:1869983.1870015} 
leveraged probe requests to show past behaviours of 
users. Similarly, Cheng \etal\cite{6415713} employed the spatial temporal information 
of probe requests to reveal the underlying social 
relationships between a small sample of users.
Barbera \etal\cite{Barbera:2013:SCU:2504730.2504742} focused on building 
a snapshot of thousands of users involving in a large scale event.

In addition, some companies have launched WiFi-based software products for e.g. 
retail shopping behavior estimation.  Such products have been 
criticized by some to be highly error-prone and their results unstable 
\cite{wificritique}. 
Indeed,  WiFi data are challenging to use as the  sending frequency of probe 
requests is highly variable,  depending 
on  the mobile station's operating system version as well as on 
its power management state (awake or 
sleeping)\cite{Musa:2012:TUS:2426656.2426685, 
Wang:2013:FSM:2500727.2500729,Desmond:2008:IUD:1352533.1352542}.
Furthermore, the observing system or AP may fail to record some probe requests because of packet loss 
during transmission or limited capacity of handling concurrent packets. Lastly, 
some MAC addresses are randomised (spoofed) in order to protect 
users' privacy. It is thus very important that the applications and the methods developed to use WiFi data be robust to the above sources of noise.

While the issues raised above make WiFi-based pedestrian counting very challenging, a new area in which WiFi analytics have promise for potentially delivering great value is the real-time  monitoring  of public transport systems. 
Indeed, we shall show that some key performance indicators and real-time operational information needed for monitoring public transport systems can be obtained with remarkably high accuracy.
 One such example is determining the true (as opposed to the published) timetable of a public transport system in real-time. In most cases, there are no other publicly-available sources for this information. Yet, it is indispensable for the monitoring of public transport service levels. Even more importantly, this information, if accurate and temporally fine-grained, can be used for very low-latency incident detection.

In this work, we propose using spectral clustering based on trajectories derived from  WiFi traces of users of a public transport system to infer the true timetable and  two key performance indicators of the transport service, namely public transport vehicle headway and station dwell time. By detecting anomalous dwell times or headways, we demonstrate that a fast and accurate real-time incident-detection procedure can be obtained. The method we introduce makes use of the advantages of the high-frequency WiFi data, which provides very low-latency, universally-accessible information, while minimizing the impact of the noise in the data.

The problem of estimating or predicting the true timetable and hence delays of 
public transport services has been addressed in a number of research papers. For 
the most part, however, these papers assume access to real-time actual train 
locations from e.g. the train's signaling system, or they use simulated data, or even cell phone data (voice call, SMS). In 
practice, though, only the train operator has access to the actual train 
locations and in general that data 
are part of a closed proprietary  system  which cannot be exposed in real-time 
for use in data analytics. In a series of papers, Kecman and Goverde
use this type of data, referred to as track occupation messages, for a number of 
applications such as train position prediction, route conflict identification,  
station dwell time prediction, etc. \cite{kecman2013online, kecman2013process, 
kecman2015online, kecman2015predictive}. Similarly, \cite{Martin2016} proposes a 
method to predict real-time train movements. 
While track occupation data are very noisy, as discussed by the authors in 
\cite{kecman2013process}, it is nonetheless a relatively reliable source of 
real-time train positions. Outside the train operator itself, however, real-time 
track occupation data is not available. 

 Close in spirit to our work is the work of
\cite{HORN201567}, who derive regional train timetables using large-scale cell phone data. They are able to  deduce a published timetable by detecting  bursts in transitions of cell phone subscribers.   The authors  report a precision of 85\% within  $\pm 5$ minutes, but with a recall rate of only 49\%. Specifically, they show that cell phone data when geo-localized against regional train stations exhibit the bursty pattern that corresponds precisely to the timetable of the regional train, and contrast that to the pattern of transitions of users geo-localized against motorway junctions, which do not show any particular pattern on the same scale. Their method works quite simply by identifying user transitions between areas within a 1km radius of a regional train station, aggregating the transitions for each station, then detecting the bursts, and associating the time of the maximum value of the burst to the train departure time. While the method works reasonably well for the highly separated regional train lines, it would not work well on an urban metro system. Indeed, the cell phone signals received by the operator are infrequent in time and have low spatial accuracy. In addition, while cell phone records can be processed by the telecommunications operator, they are not available to the general public.

Our contribution is to define a means for using passive, universally-accessible WiFi data to estimate 
train movements in real time.  The method is based on spectral clustering of  
derived journeys of individual travellers in the public transport network. Our 
spectral clustering-based approach permits obtaining directly an estimate of 
each train, in space and time, without requiring apriori knowledge of the 
number of trains in the network. Outlier detection is critical as the WiFi data 
are inherently noisy; we propose a two-part outlier detection approach to 
handle the two main types of outliers associated with using WiFi data.
The method is evaluated on data from one line of a metropolitan train (subway) 
system in a large city. 
It is  shown that the results are highly accurate both 
during light flow conditions as well as during peak travel times in which train 
frequency is high and cluster separation could potentially be problematic. In 
addition, the method is tested on data obtained during a train disruption; it is 
shown that it can be effectively used for incident detection. Numerical 
evaluation of the method is provided against two different baselines.

The organisation of the paper is as follows. In the next section, we present 
the main characteristics of the WiFi data and a data aggregation method to 
enable use of our proposed spectral clustering approach. This includes the 
definition of a certain type of train journey for each passenger having a WiFi 
enabled device.  Section 3 presents a baseline method using station-specific 
clustering as well as our proposed spectral clustering method and 
the outlier reduction strategies developed. In Section 
4 we present the evaluation of our method on WiFi data from two weeks of both 
peak and off-peak travel as well as during a train disruption. Section 5 
concludes with a few avenues for further work in this area.

\section{Preliminaries}

 WiFi data are obtained as a sequence of records for each device and consist 
of at a minimum of fields including an anonymous identifier of the device (the 
MAC address) as well as a timestamp and location (or in some cases a signal 
strength value). The update frequency of the data may be every time a probe 
request is sent by a device such as when a WiFi sniffer is used, or may be 
pre-aggregated (by an AP) to the level of e.g.\ each second. Recall that 
devices send probe requests at highly variable frequencies across the population 
of all devices. In some cases, one device may be recorded every second, for 
other devices the frequency of transmissions may be considerably less. 

In order to make use of Wifi data then, it is not possible 
assume that lack of an observation of a particular device implies that it is no 
longer present. Rather, the aggregation method employed must be robust to the 
highly-variable frequency of observations across devices. To this end, we 
introduce the following notion of a \emph{physical journey} and a
$\tau$-\emph{journey}.

\begin{definition}[physical journey]
\label{def:mrt:physical-journey}
    Given a MAC address $m$ and $\mathcal{R}_m:=\{(v_i, t_i)\}_{i = 1}^n$ $n$ 
records associated with $m$. A 
subset $\mathcal{R}_m'\subset\mathcal{R}_m$ is called a \emph{physical journey}
if
\begin{enumerate}
    \item $\mathcal{R}_m'$ belongs to a single train;
    \item if $\mathcal{R}_m'' \subset{R}_m$ satisfies
$1$ and $\mathcal{R}_m' \subset \mathcal{R}_m''$, then 
$\mathcal{R}_m' = \mathcal{R}_m''$.
\end{enumerate}
\end{definition}
Generally, no information about the physical journey is 
available. Therefore we introduce the notion of a $\tau$-\emph{journey}, or 
simply \emph{journey}, to approximate physical journey.

\begin{definition}[$\tau$-journey]
\label{def:mrt:journey}
    Given a MAC address $m$ and $\mathcal{R}_m:=\{(v_i, t_i)\}_{i = 1}^n$ $n$ 
records associated with $m$ such that $t_1 \leq t_2 \leq \ldots \leq t_n$. A 
subset $\mathcal{R}_m'\subset\mathcal{R}_m$ is called a $\tau$-journey if
    \begin{enumerate}
     \item (\emph{temporal continuity}) $r_i, r_j \in \mathcal{R}_m'$ s.t.\ 
$i < j$, 
we have $r_k \in \mathcal{R}_m'$ for all $r_k \in \mathcal{R}_m$ such that $i < 
k < j$;
     \item (\emph{intra-station continuity}) 
$\displaystyle\max_{r_i,r_j\in\mathcal{R}_m', v_i = v_j}|t_i - t_j| \leq 
\tau_1$;
     \item (\emph{inter-station continuity}) 
$\displaystyle\max_{r_i,r_j\in\mathcal{R}_m', v_i \neq v_j}|t_i - t_j| \leq 
\tau_2$;
     \item (\emph{monotonicity}) $\mathrm{sgn}\left((t_i - t_j)(v_i - v_j)\right)$ 
is constant for all $r_i, r_j\in\mathcal{R}_m'$;
     \item (\emph{maximality}) if $\mathcal{R}_m'' \subset{R}_m$ satisfying 
$1$, $2$, $3$ and $4$ and $\mathcal{R}_m' \subset \mathcal{R}_m''$, then 
$\mathcal{R}_m' = \mathcal{R}_m''$.
    \end{enumerate}
\end{definition}

The value of $\tau_1$ is  the average time between two \emph{receives} 
of probe requests of the same device. 
 Wang \etal\cite{Wang:2013:FSM:2500727.2500729} proposed values of
around one or two minutes as the time between two \emph{sends} of probe 
requests. Our models are more 
robust to the case where several actual journeys are identified as a single one 
than the case where one actual journeys is identified as several ones. 
Therefore, we choose  a  more conservative $\tau_1 = 8\textrm{min}$ as 
suggested in 
\cite{Musa:2012:TUS:2426656.2426685}. 

The value of $\tau_2$, on the other hand, depends  on the
train itself. 
Since trains may experience delays 
we choose a larger value such as $\tau_2=30$min. In general, this value should be chosen so that 
 it is  unlikely that two records of a  journey 
are more than $\tau_2$ minutes apart.

\begin{figure}[htbp!]
    \centering
    \includegraphics[width = 0.9\textwidth]{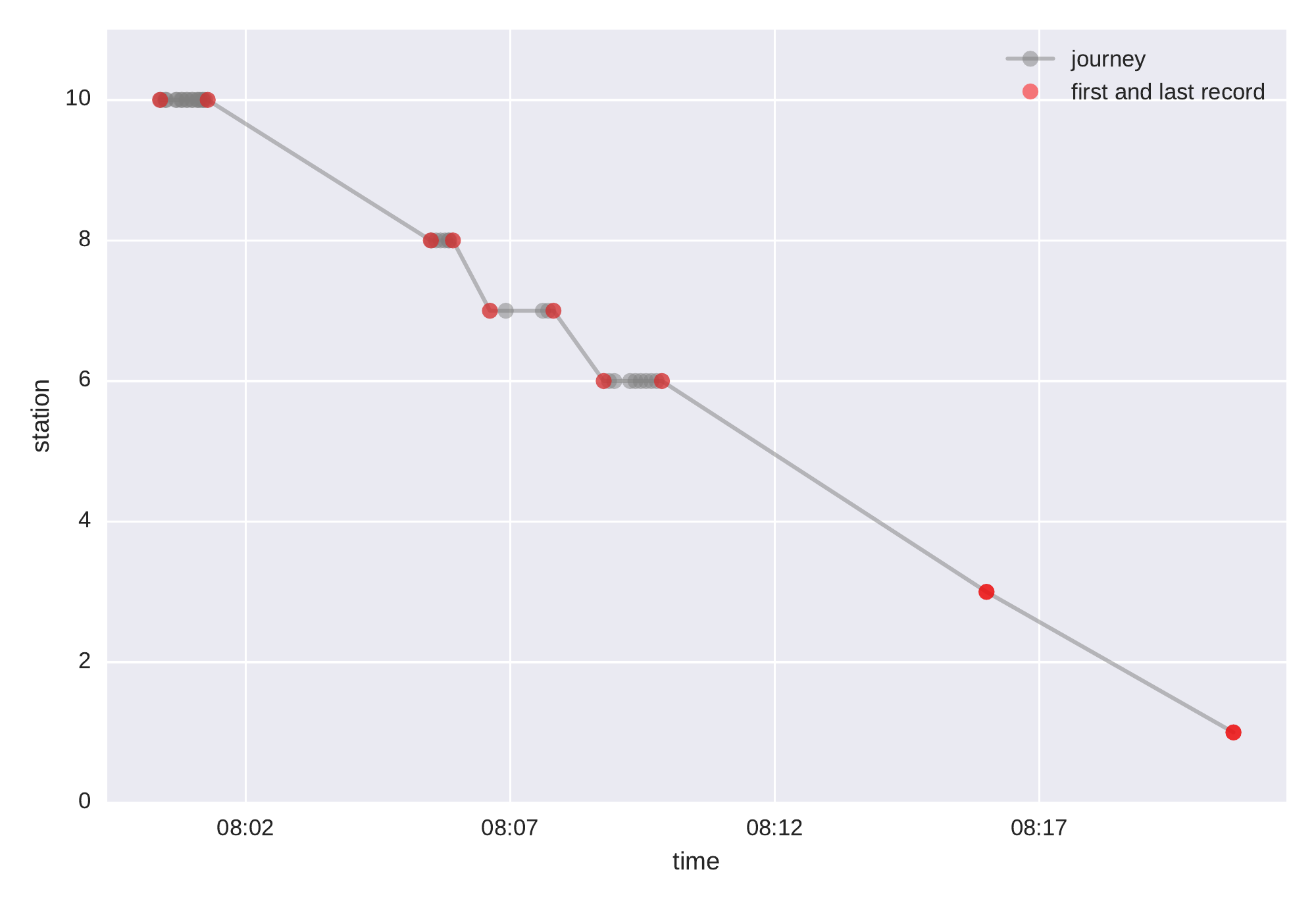}
    \caption{Example of a journey through $10$ train stations with the first 
and the last WiFi
record observed each station coloured in red.}
    \label{fig:mrt:example_journey}
\end{figure}

Figure \ref{fig:mrt:example_journey} illustrates a derived journey, as defined 
in 
Definition \ref{def:mrt:journey}.
 We see in particular that the device 
was recorded by APs not only when the owner was on his boarding and alighting station platforms
but also when he/she was in the train itself, that is, when the train was in the intermediate stations during the journey.
Observe further that, for a 
given journey, at each station, only the first and the 
last record at each station provide new information. 
Eliminating intermediate records aids in reduction of the WiFi data size by 
several orders of magnitude.

\begin{figure}[htbp!]
    \centering
    \includegraphics[width = 0.8\textwidth]{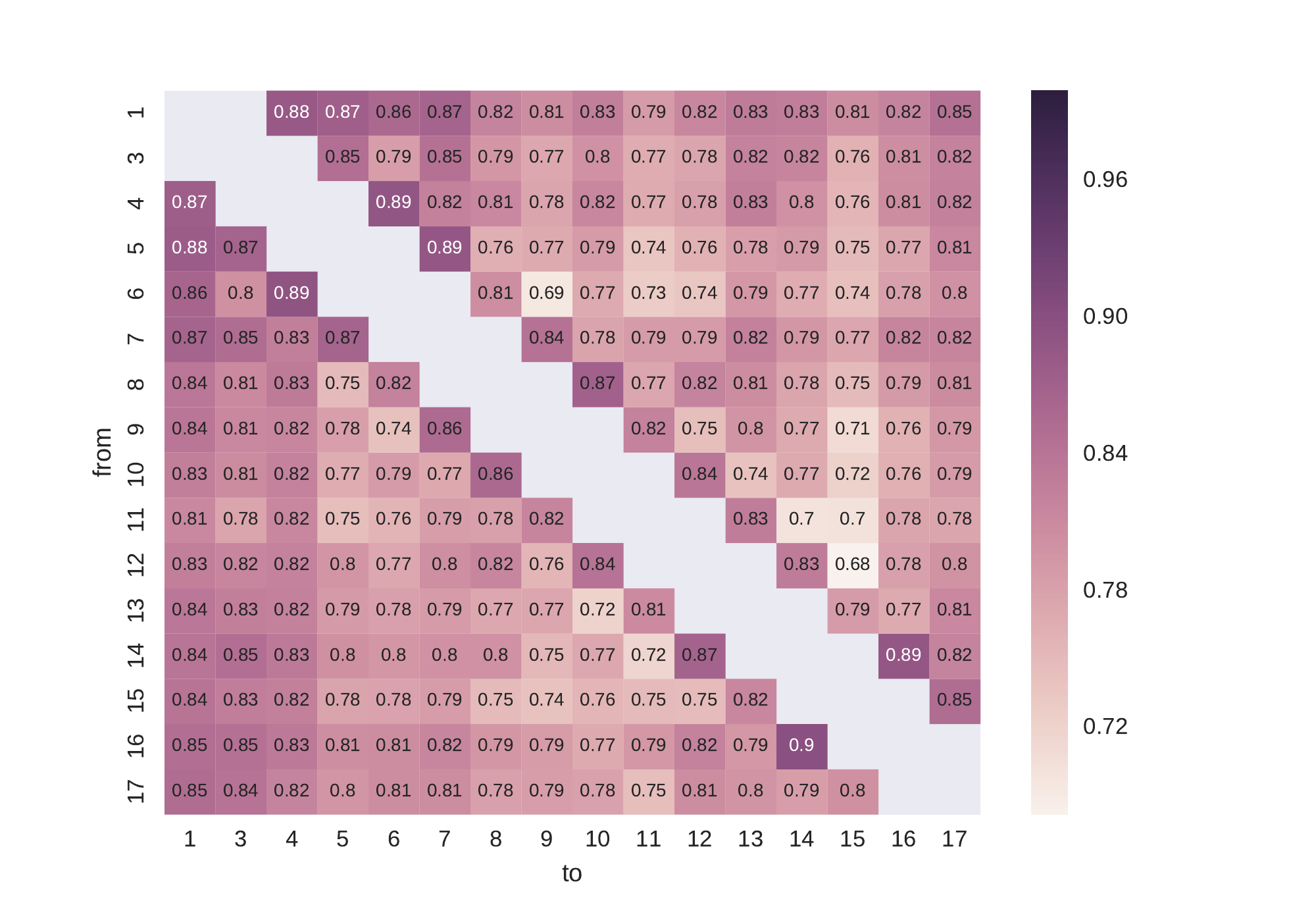}
    \caption{Heat map of average ratio of number of stations where a 
device is recorded}
    \label{fig:mrt:heatmap}
\end{figure}

Figure \ref{fig:mrt:heatmap} shows the average ratio 
of stations where 
a device is recorded with respect to stations at which the device actually 
passes, \emph{i.e.}\  
    \[
    P_{i,j} = \mathbb{E}_{\textrm{device } k 
        \textrm{ from } i \textrm{ to } j}\left[\frac{\#\{\textrm{stations 
where } k \textrm{ is recorded}\}}
        {\#\{\textrm{stations which } k \textrm{ passed}\}}\right]
    \] 
for each journey from station $i$ to station $j$\footnote{Note that the $3$ main diagonals 
are 
masked 
since all entries are ones.}.  Empirically in this example, one observes that a device is likely to be 
recorded at roughly $60\%$ of stations it passes through. 
    
We are interested in estimating the following two quantities:
\begin{definition}[Dwell time, Headway]
   For a given train at a given station, we call
    \begin{itemize}
     \item the \emph{dwell time} the time that the train \emph{stays} at the 
station;
     \item the \emph{headway} is the time difference before the departure of 
this train and the arrival of next train.
    \end{itemize}
\end{definition}

\begin{figure}[htbp!]
    \centering
    \begin{subfigure}[t]{0.45\textwidth}
        \includegraphics[width = 0.9\textwidth]{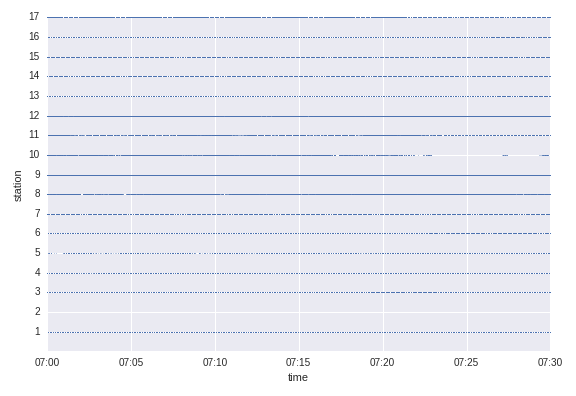}
        \caption{Raw data forming a continuum along the $x$ axis}
        \label{fig:mrt:raw}
    \end{subfigure}
    \begin{subfigure}[t]{0.45\textwidth}
        \includegraphics[width = 0.9\textwidth]{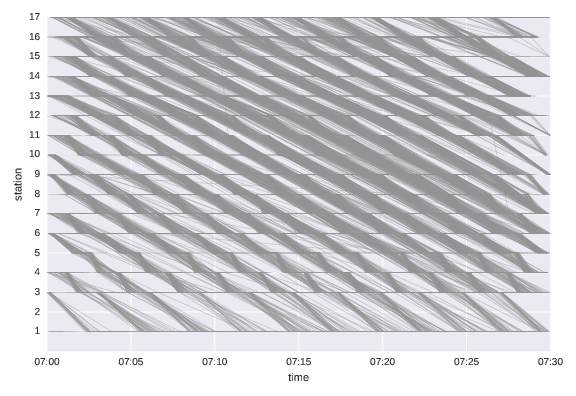}
        \caption{End-to-end estimated journeys. Note the high degree of noise at the boarding and alighting 
stations}
        \label{fig:mrt:journey-raw}
    \end{subfigure}
    \begin{subfigure}[t]{0.45\textwidth}
        \includegraphics[width = 0.9\textwidth]{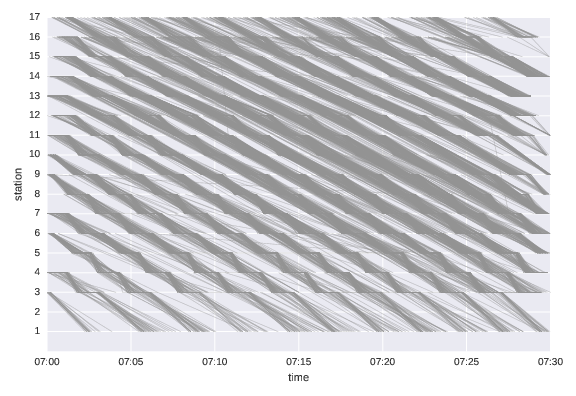}
        \caption{Estimated journeys where  the first WiFi record at the commuter's \emph{boarding} station 
and 
the last record at the \emph{alighting}
station were removed.}
        \label{fig:mrt:journey-inner}
    \end{subfigure}
    \begin{subfigure}[t]{0.45\textwidth}
        \includegraphics[width = 0.9\textwidth]{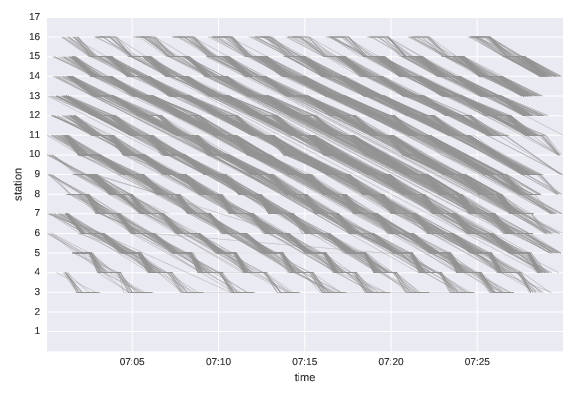}
        \caption{Estimated journeys in which all records at \emph{boarding}
and \emph{alighting} stations were removed.}
        \label{fig:mrt:journey-sans}
    \end{subfigure}
    \caption{ Data cleaning strategies for estimating individual journeys. }
    \label{fig:mrt:strategies}
\end{figure}

Deriving  traveller  journeys is critical for the spectral clustering method to 
perform well. Indeed, the WiFi data at each station, as shown in 
Figure \ref{fig:mrt:raw},   form almost a continuum 
along the  
time axis at each station. 
Figures (\ref{fig:mrt:journey-raw}), (\ref{fig:mrt:journey-inner}) 
and (\ref{fig:mrt:journey-sans}) show data cleaning strategies at three 
different 
levels: 
\begin{itemize}
 \item Level 1: Figure (\ref{fig:mrt:journey-raw}) shows \emph{all} estimated 
journeys according to Definition 1. Note in particular the considerable noise 
at the boarding and alighting stations of each journey. This is an intrinsic error associated with 
 using WiFi data as 
passengers are recorded upon entry to the station and  
wait different amounts of time on the platform before boarding and  after  
alighting; 
 \item Level 2: in Figure (\ref{fig:mrt:journey-inner}), 
 only the last recorded observation of each device at the boarding station (and 
the first observation at the alighting station) and observations of all 
``internal'' stations are used in the trajectory estimation.
 \item Level 3: in Figure 
(\ref{fig:mrt:journey-sans}) records of boarding and alighting stations are 
removed. The resulting journeys are then clearly identifiable.
\end{itemize}

Using level 3 data cleaning, the first and last station
of a train are lost. 
Thus, a  hybrid strategy is employed:
 level 3 serves to identify trains and associate journeys to them, 
and  level 2 serves to determine the train timetable. That is, 
the origin station on the line is added back once the train cluster is identified.

\section{Methods}
\label{sec:mrt:stationwise}

Our  baseline method for determining the real-time train timetable from WiFi records involves performing  1-D clustering of 
timestamps at each station and then identify clusters at different stations 
belong to the same train by majority vote. The method is summarized in Algorithm \ref{alg:mrt:stationwise}  in the Appendix. While we choose to use 
DBSCAN\cite{473950} in the scope of this paper, other clustering methods could 
also be used instead.

\begin{figure}[htbp!]
    \centering
    \includegraphics[width = 0.7\textwidth]{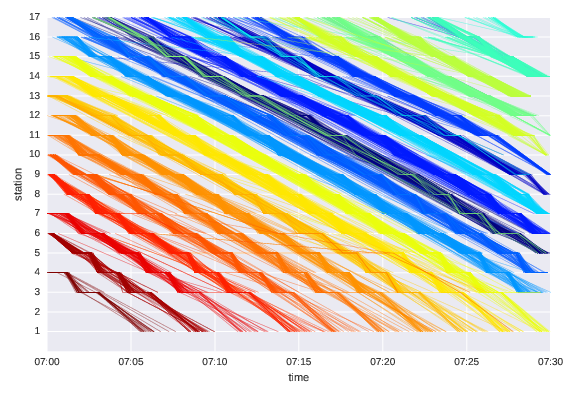}
    \caption{Clusters of figure \ref{def:mrt:journey} by the baseline method 
        with $\epsilon = 6$ and min\_samples $= 10$ for DBSCAN. 
        Each colour represents a cluster label.}
    \label{fig:mrt:staionwise-cluters}
\end{figure}

Figure \ref{fig:mrt:staionwise-cluters} shows the result of the baseline 
method. 
Station-specific clustering, while simple and straightforward,   
 can
fail to separate 
two trains at one station, and as a result all subsequent stations along the 
line will be influenced.
Consider the following  example\footnote{See also figure \ref{fig:mrt:raw}, 
    around 07h25 at station $9$.}: a line with three 
stations $A, B$ and $C$ and a train from $A$ to $C$:
\[
    A \longrightarrow B \longrightarrow C
\]
Suppose further that at station $B$ no passengers having boarded at $A$ are  
observed at $B$, but all passengers boarding at $A$ and $B$ are recognised 
at $C$. This type of anomaly can occur when e.g. a WiFi router is saturated.
 Algorithm \ref{alg:mrt:stationwise} will consider there to be
a 
new train at $B$, and errors may be introduced at station  $C$, since some of the MAC 
addresses have been seen at $A$ whereas others have been seen at $B$. Spectral clustering addresses this problem as it considers the trajectories of travelers and can thus readily interpolate between stations where observations are missing. 
Though Figure \ref{fig:mrt:staionwise-cluters}  appears correct, we shall see later that the baseline method  fails to identify two trains. 

Spectral clustering when applied to the estimated journeys can be much more accurate
than the baseline method. In addition the method is applicable 
to both normal and incident 
situations. Spectral clustering requires the definition of a similarity matrix, and uses the eigenvalues of that matrix for dimensionality reduction in the definition of the clusters.
An intuitive vectorization to use  is to embed a journey into 
$\overline{\mathbb{R}^{s}} = \mathbb{R}^{s} \cup \{\infty\}$ by identifying a 
journey with 
\begin{equation}
    \bm{t} = (t_1, t_3, t_4, \ldots, t_{s})
    \label{eqn:mrt:vectorization}
\end{equation}
where $t_k \in \mathbb{R}_+ \cup \{\infty\}$ is the mean of 
timestamps\footnote{Strictly speaking, since for a given journey, only two 
extremity timestamps at each station are kept, the \emph{mean} timestamp here 
corresponds to the midpoint of timestamps of original data.} at 
stations $k=1\ldots s$ with $t_k = \infty$ if the device is not recorded at $k$. 
In order to assess pairwise similarity, we require the following definitions:

\begin{definition}[$l^0$ norm]
    Let $\bm{t}$ be a point as in (\ref{eqn:mrt:vectorization}), the 
$l^0$ norm of $\bm{t}$ is defined as the number of non-infinite entries of 
$\bm{t}$:
    \[
        \|\bm{v}\|_1 = \sum_{i = 1, 3, 4, \ldots, 
s}{\mathds{1}_{t_i = \infty}}.
    \]
\end{definition}

\begin{definition}[$l^{\infty}$ norm]
\label{def:mrt:l-infinite}
     Let $\bm{t}$ be a point as in (\ref{eqn:mrt:vectorization}), the 
$l^{\infty}$ 
norm of $\bm{t}$ is defined as the maximum absolute value of non-infinite 
entries
\[
    \|\bm{t}\|_{\infty} = \left\{
                                 \begin{array}{rcl}
                                    \displaystyle\max_{\substack{i = 1, 3,4, 
\ldots, s\\t_i\neq \infty}}{|t_i|}&,&\textrm{if 
 }\|\bm{v}\|_0\neq 0,\\
                                    \infty&,&\textrm{otherwise.}
                                 \end{array}
                            \right.
\]
\end{definition}

\begin{definition}[difference]
    For $\bm{t}_1,\bm{t}_2$ two points as in (\ref{eqn:mrt:vectorization}), 
The difference of $\bm{t}_1,\bm{t}_2$ is defined as
    \[
        \bm{t}_1 - \bm{t}_2 = (t_{1,1} - t_{2,1}, t_{1,3} - t_{2,3}, t_{1,4} - 
t_{2,4}, \ldots, t_{1,s} - t_{2,s})
    \]
with $\infty - * = \infty, * - \infty = \infty$ and $\infty - \infty = \infty$.
\end{definition}

\begin{definition}[pairwise similarity]
\label{def:mrt:similarity}
    For $\bm{t}_1,\bm{t}_2$ two points as in (\ref{eqn:mrt:vectorization}), we 
define the following two pairwise similarity metrics:
\begin{equation}
    \mathrm{sim}_\mathrm{soft}(\bm{t}_1, \bm{t}_2) = \|\bm{t}_1 - \bm{t}_2\|_0
\exp\left(-\frac{\|\bm{t}_1 - \bm{t}_2\|_{\infty}^2}{2\sigma^2}\right).
\label{eqn:mrt:soft}
\end{equation}
\begin{equation}
    \mathrm{sim}_\mathrm{hard}(\bm{t}_1, \bm{t}_2) = \|\bm{t}_1 - \bm{t}_2\|_0
\mathds{1}_{\|\bm{t}_1 - \bm{t}_2\|_{\infty} \leq \tau}.
\label{eqn:mrt:hard}
\end{equation}
\end{definition}
\begin{remark}
    In (\ref{eqn:mrt:soft}) and (\ref{eqn:mrt:hard}), the $l^0$ term quantifies 
    the \emph{spatial} similarity, \emph{i.e.}\ number of stations where both 
    journeys are recorded; the $l^{\infty}$ term quantify the \emph{temporal} 
    similarity, \emph{i.e.}\ maximum time difference at stations where both 
    journeys are recorded.
\end{remark}

\begin{definition}[similarity graph]
\label{def:mrt:graph}
Given $N$ points $V = \{\bm{t}_i\}_{i=1}^N$ as in 
(\ref{eqn:mrt:vectorization}), the similarity graph $\mathcal{G} = 
(V, E)$ is such that an edge $e_{ij} = (\bm{t}_i, 
\bm{t}_j)$ of 
weight $\mathrm{sim}(\bm{t}_1, \bm{t}_2)$ exists if 
$\emph{sim}(\bm{t}_1, \bm{t}_2) > 0$. 
\end{definition}

\begin{figure}[htbp!]
    \centering
    \includegraphics[width = 
0.7\textwidth]{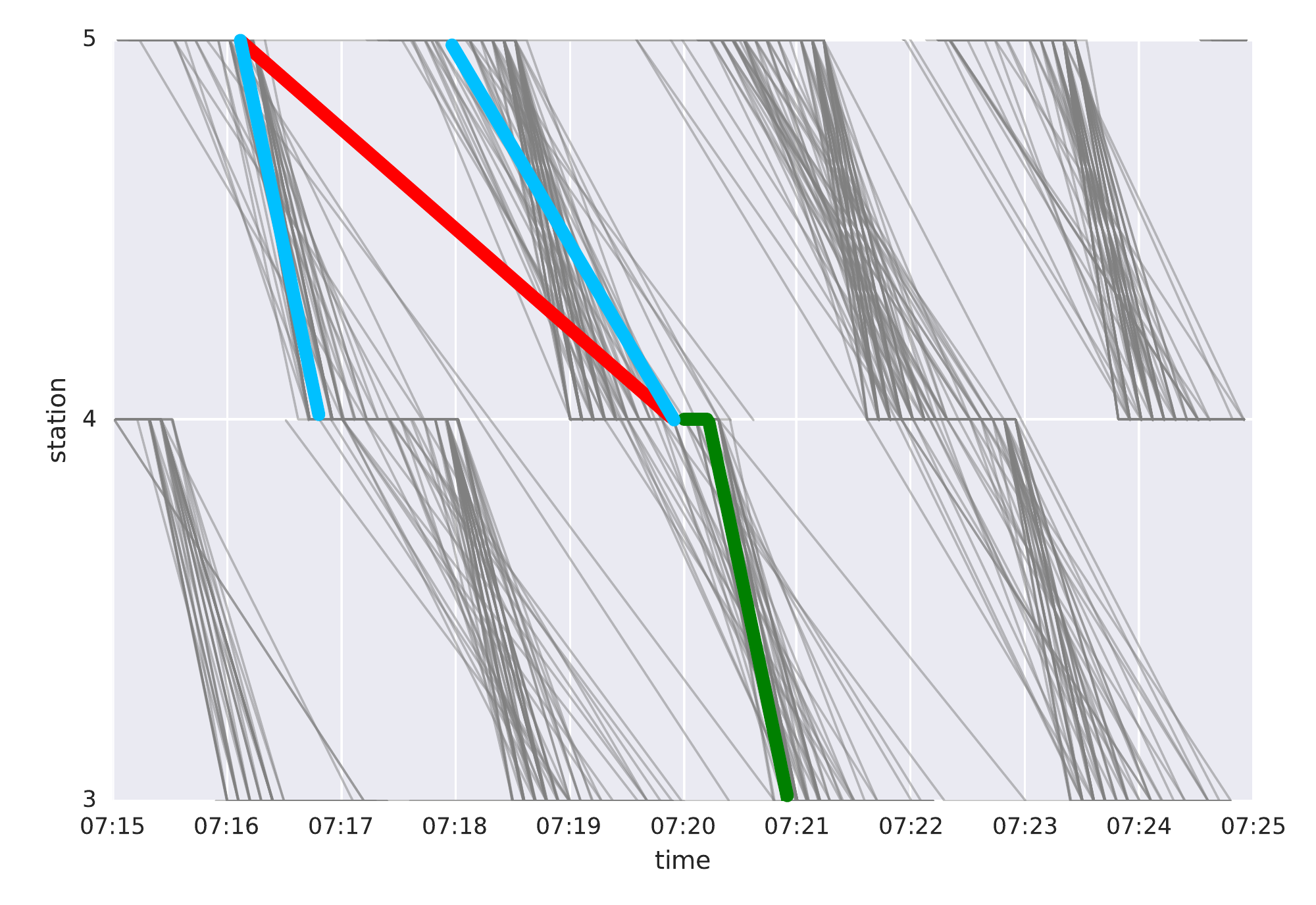}
    \caption{Behaviour of Definition \ref{def:mrt:similarity} on mis-identified 
        journeys. The red journey is mis-identified and is dissimilar to the 
two blue 
        journeys. However, it is similar to the green journey since their 
        timestamps at 
        their only overlapping station are similar.}
    \label{fig:mrt:mis-identified}
\end{figure}

\begin{figure}[htbp!]
    \centering
    \includegraphics[width = 0.3\textwidth]{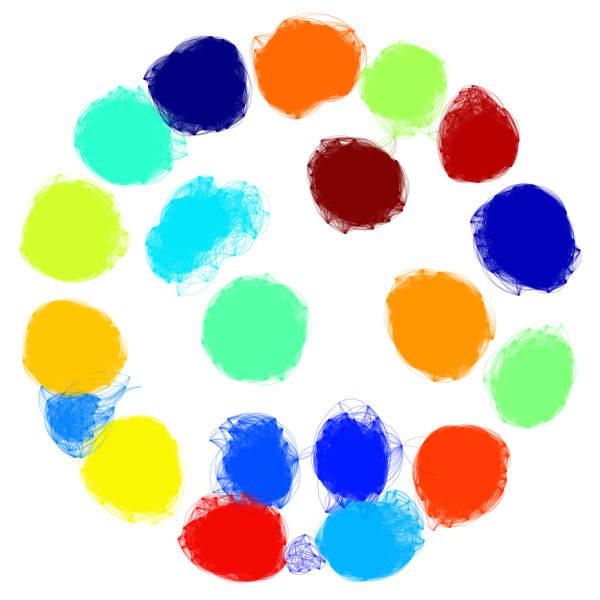}
    \caption{Similarity graph of data shown in Figure 
\ref{fig:mrt:journey-sans}.}
    \label{fig:mrt:graph}
\end{figure}

Figure \ref{fig:mrt:mis-identified} illustrates Definition 
\ref{def:mrt:similarity} on  a mis-identified journey  (in 
red).  The blue journeys are dissimilar to the red journey by the 
exponential  in (\ref{eqn:mrt:soft}) and the indicator function in 
(\ref{eqn:mrt:hard}).  
 The  mis-identified journeys thus have  very low degree in the 
similarity graph, and thus the two  are unlikely to be  in a 
single cluster using spectral clustering. Figure \ref{fig:mrt:graph} shows 
the similarity graph of the estimated journeys illustrated in Figure 
\ref{fig:mrt:journey-sans}.  One  observes $21$ well defined clusters with 
very few intra-cluster links.

Let $\mathcal{G} = (V, 
E)$ be an undirected graph with $n$ vertices and $W$ its 
weighted adjacency matrix. For a vertex $v_i\in V$, the \emph{degree} 
of $v_i$ is 
\[
 d_{i} = \sum_{j = 1}^{n}w_{i,j}, 
\]
and the \emph{degree matrix} $D$ is 
defined as 
\[
    D = \begin{pmatrix}
            d_1&&&&\\
            &d_2&&&\\
            &&\ddots&&\\
            &&&d_{n-1}&\\
            &&&&d_{n}
        \end{pmatrix}.
\]
Set $A\subset V$ is \emph{connected} if any two vertices in $A$ 
can be joined by a path in $A$.
Define $i\in A$ to be $\{i|v_i\in A\}$.
Then, we denote $|A|$ the number of vertices in $A$ 
and $\textrm{vol}(A) = \sum_{i\in A}d_i$. 
Furthermore, 
for two sets $A, B\subset V$, we define
\[
    W(A, B) = \sum_{i\in A, j\in B}w_{i,j}.
\]
Denote $\overline{A} = V - A$ the complement of $A$. We say $A$ is a 
\emph{connected component} if $A$ is connected and $W(A, \overline{A}) = 0$.  
Non empty sets $A_1, A_2, \ldots A_k \subset V$ form a 
\emph{partition} of $\mathcal{G}$ if $A_i\cap A_j = \varnothing$ for all $i\neq 
j$ and $\displaystyle\cup_{i = 1}^{k}A_i = V$.
Denote $L = D-W$ the \emph{non-normalised Laplacian} and $L_{\textrm{rw}} = 
D^{-1}L = I - D^{-1}W$ the \emph{normalised Laplacian} of $\mathcal{G}$. We 
have the 
following result\cite{vonLuxburg2007}:
\begin{proposition}
    The normalised Laplacian $L_{\textrm{rw}}$ satisfies:
    \begin{itemize}
     \item $L_{\textrm{rw}}$ is positive semi-definite;
     \item $L_{\textrm{rw}}$ has $n$ non-negative real-valued eigenvalues $0 
= \lambda_1 \leq 
\lambda_2 \leq \ldots \leq \lambda_n$;
    \item the multiplicity $k$ of eigenvalue $0$ equals the number of connected 
components $A_1, A_2, \ldots, A_k$ of $G$.
    \end{itemize}
\end{proposition}

Given $k$ a number and $A_1, A_2, \ldots, A_k$ a \emph{partition} of $G$, 
the \emph{normalised 
Ncut} is defined as 
\[
    \textrm{Ncut}(A_1, A_2, \ldots, A_k) = \frac{1}{2}\sum_{i = 
1}^{k}\frac{W(A_i, \overline{A_i})}{\textrm{vol}(A_i)}.
\]
The minimisation of \emph{Ncut} clearly solves the clustering 
problem. Further, it is worth noting that \emph{Ncut} is robust to noise 
as it avoids  small (and singleton) clusters.  
 Solving \emph{Ncut} however  is NP-hard. Thus we use normalised spectral 
clustering 
 (see  Algorithm \ref{alg:mrt:spectral} in the Appendix) which solves a 
relaxation of 
\emph{Ncut}.

\begin{figure}[htbp!]
    \centering
    \begin{subfigure}[b]{0.45\textwidth}
        \includegraphics[width = \textwidth]{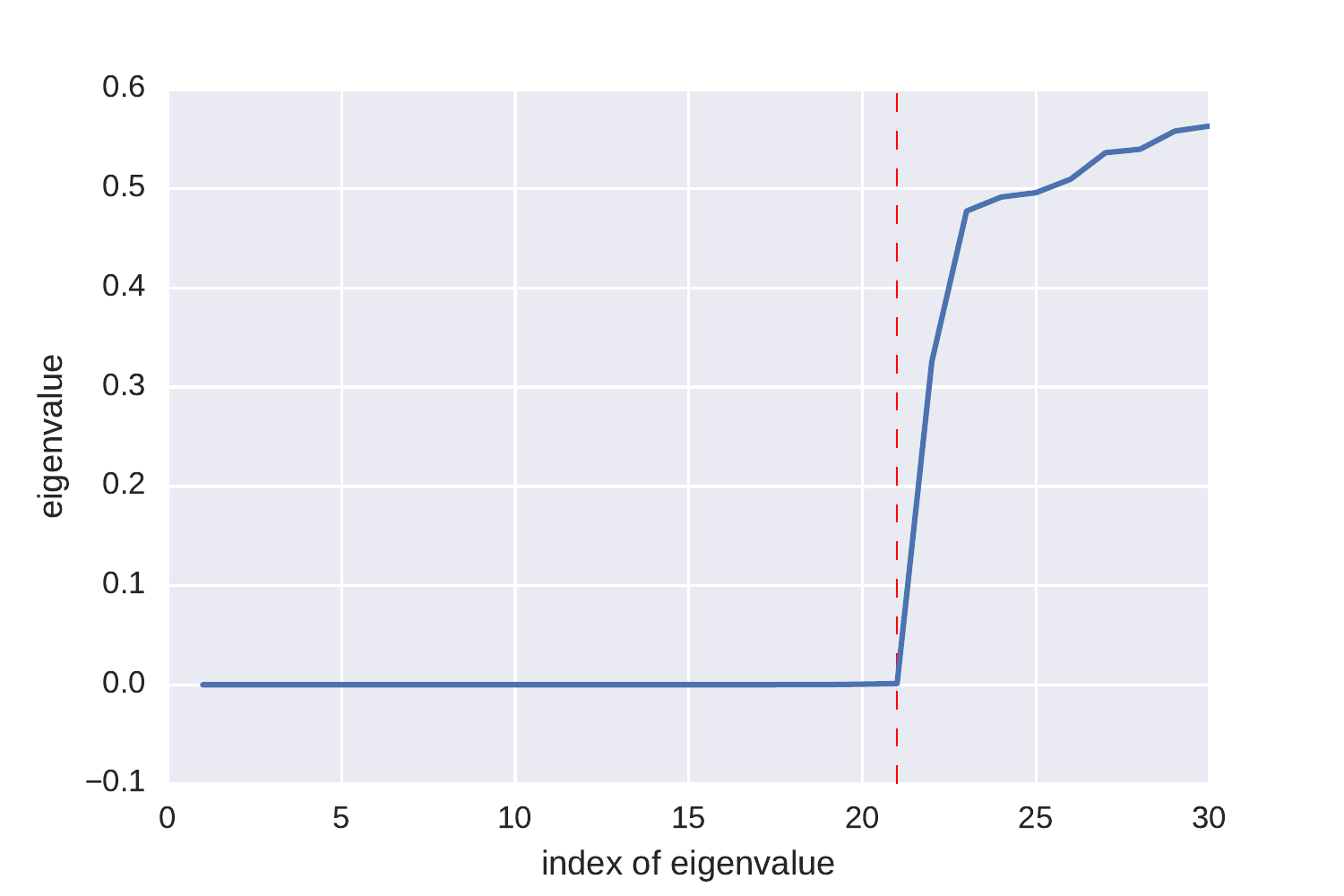}
        \caption{Hard similarity, $\tau = 30$.}
    \end{subfigure}
    \begin{subfigure}[b]{0.45\textwidth}
        \includegraphics[width = \textwidth]{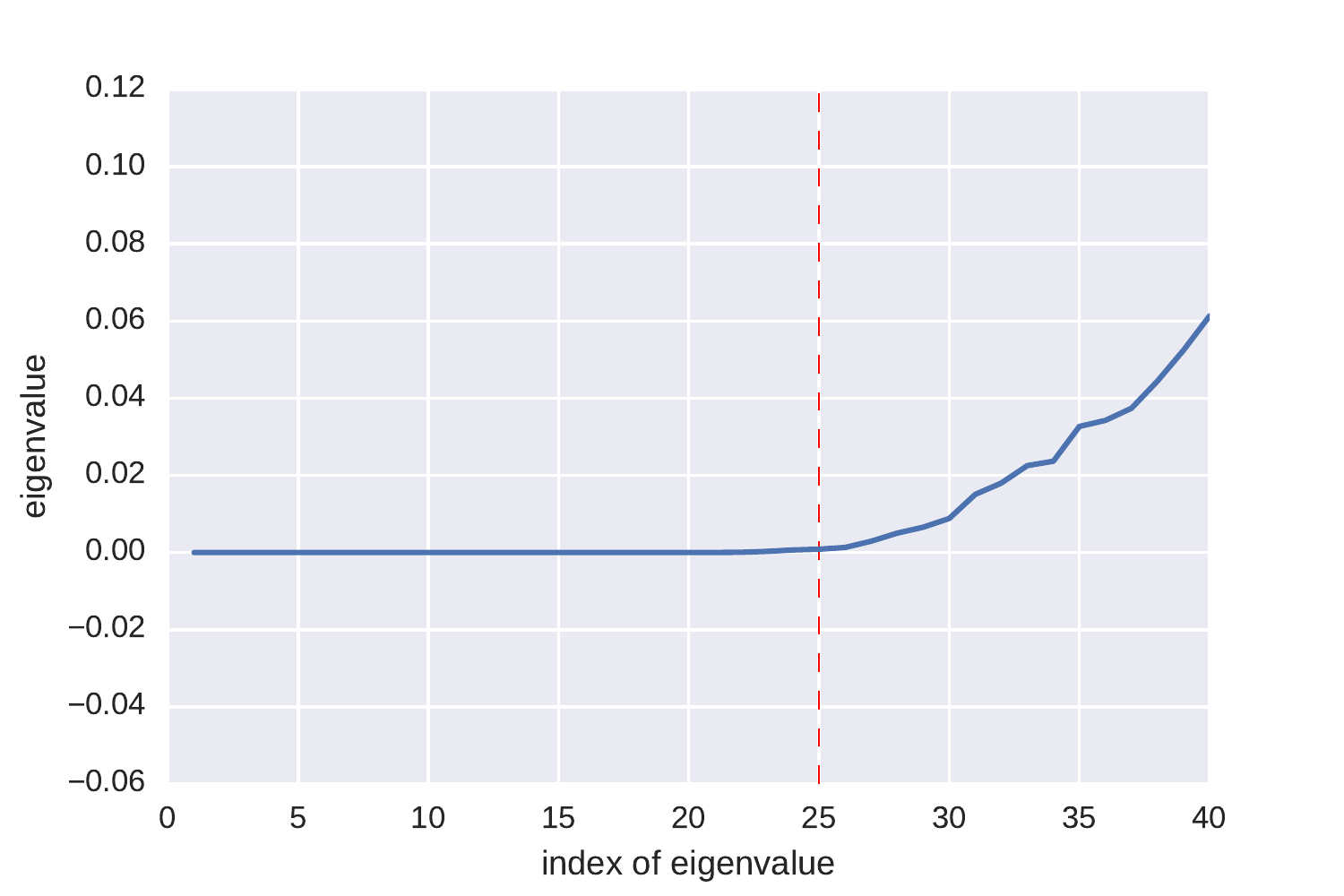}
        \caption{Soft similarity, $2\sigma^2 = 30$.}
    \end{subfigure}
    \caption{First $30$ eigenvalues of the normalised Laplacian of the graph 
associated 
        with Figure \ref{fig:mrt:journey-sans}.}
    \label{fig:mrt:eigens}
\end{figure}

In practice, it not  easy to choose the number of clusters $k$ 
\emph{apriori}. However, as we can see from Figure \ref{fig:mrt:graph}, 
the similarity graph is sparse. One can use an \emph{eigengap} 
heuristic\cite{vonLuxburg2007} to 
determine \emph{apriori} the optimal value of $k$ as a function of the 
magnitude of the 
eigenvalues.
Indeed, Figure \ref{fig:mrt:eigens} shows the first $30$ and $40$ eigenvalues 
of graph $\mathcal{G}$ associated with Figure \ref{fig:mrt:journey-sans}; 
in that example, both the  hard and soft similarity metrics give an
eigengap around $\lambda_{22}$ and 
$\lambda_{25}$. Algorithm \ref{alg:mrt:spectral} is extended to make use of the 
eigengap 
heuristic in  
Algorithm \ref{alg:mrt:spectral2}, also in the Appendix.

\begin{figure}[htbp!]
    \centering
    \begin{subfigure}[b]{0.45\textwidth}
        \includegraphics[width = \textwidth]{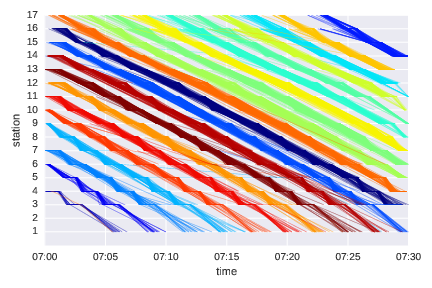}
        \caption{Hard similarity, $\tau = 30$.}
        \label{fig:mrt:clusters-hard}
    \end{subfigure}
    \begin{subfigure}[b]{0.45\textwidth}
        \includegraphics[width = \textwidth]{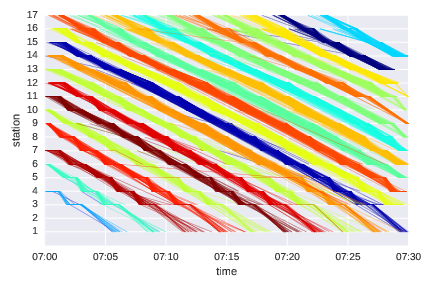}
        \caption{Soft similarity, $2\sigma^2 = 30$.}
    \end{subfigure}
    \caption{Clusters obtained from the data of Figure \ref{def:mrt:journey} 
with the spectral clustering method. 
        Each colour represents a cluster label.}
    \label{fig:mrt:spectral}
\end{figure}

\begin{table}[htbp!]
    \centering
    \begin{tabular}{ccc}
        \hline
        & soft similarity & hard similarity\\
        \hline
        matrix size & $3047 \times 3047$ & $3047 \times 3047$\\
        sparsity & $91.3\%$ & $97.1\%$\\
        execution time & $15.1$s & $12.9$s \\
        \hline
    \end{tabular}
    \caption{Comparison of the soft and hard similarity metrics on the estimated 
journeys in 
        Figure \ref{fig:mrt:journey-sans}}
    \label{tab:benchmark_soft_hard}
\end{table}

Figure \ref{fig:mrt:spectral} shows the result of spectral clustering, and 
Table \ref{tab:benchmark_soft_hard} provides a comparison of the two similarity 
measures. 
One can see that the soft similarity metric fails at times to separate  
clusters 
(See for example the cluster  at 7:00 at station 
$13$ and the cluster  at 7:00 at station $10$). 

 As with the baseline method, many outliers appear in the raw result. Since it 
is difficult to detect those outliers \emph{apriori}, we develop a set of two outlier detection methods, discussed  next.

\begin{figure}[htbp!]
    \centering
    \includegraphics[width = 0.7\textwidth]{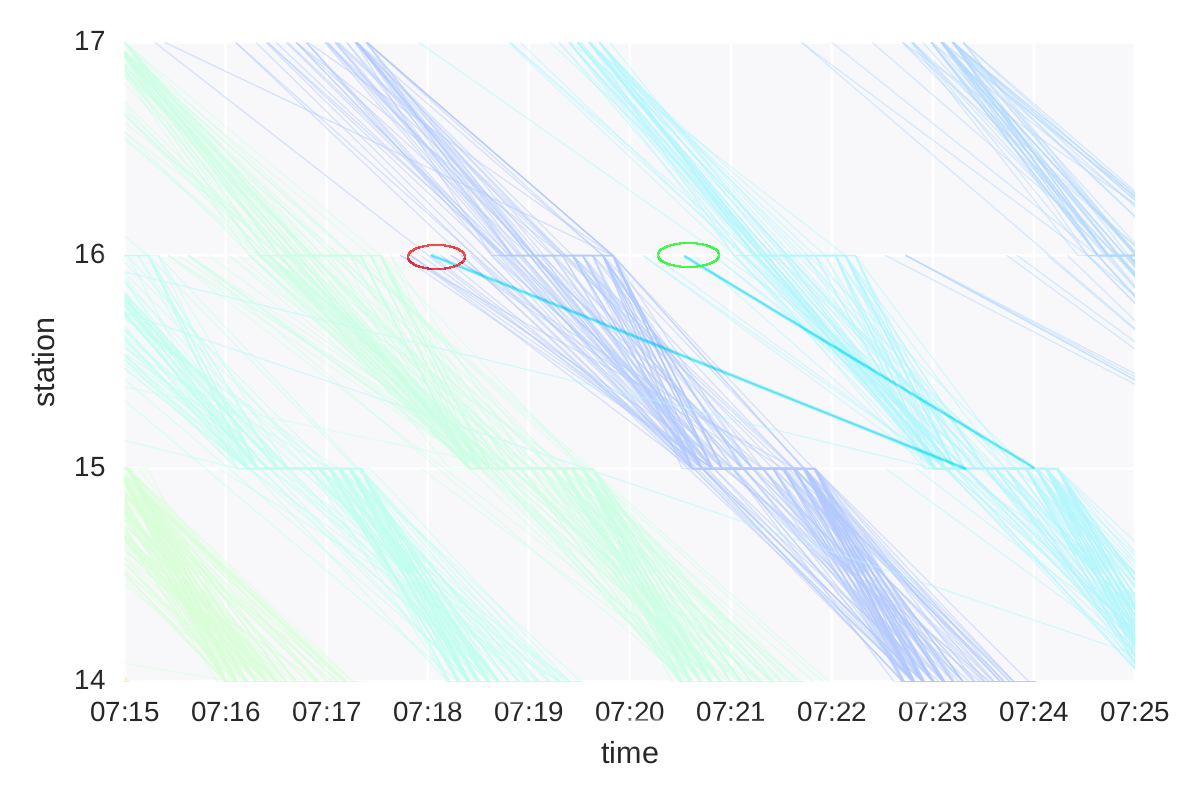}
    \caption{Two types of outliers: type 1 in red, and type 2 in green.}
    \label{fig:mrt:outliers}
\end{figure}

 Figure \ref{fig:mrt:outliers}  illustrates two types of anomalies 
in the estimated journeys arising from the use of WiFi data. The clusters are 
those of Figure \ref{fig:mrt:clusters-hard}.

 \emph{Type 1} outliers, shown in red in Figure \ref{fig:mrt:outliers}, 
occur in the journey estimation step. This may represent an instance where 
the device  owner failed to
board a train but his device did not send any further probe requests while on 
the platform at station 16. 
The record highlighted in 
red  should thus have been flagged as a separate 
journey.

 \emph{Type 2} outliers, shown in green, can be considered intrinsic WiFi 
outliers as they are due to the unpredictable sending of probe requests. In this 
example, the last record at the 
boarding station or the first record at the alighting station does not match 
the 
train departure or arrival. Such WiFi records should be dropped rather than 
integrated into any journey.

\begin{comment}
\begin{figure}[htbp!]
        \centering
        \includegraphics[width = 0.4\textwidth]{imgs/toy_graph.png}
        \caption{Similarity graph of data shown in 
\ref{fig:mrt:journey-sans} coloured by cluster labels}
    \label{fig:mrt:graph}
\end{figure}
\end{comment}

Figure \ref{fig:mrt:graph} shows the similarity graph of the estimated journeys 
in Figure 
\ref{fig:mrt:journey-sans} with vertices coloured according to cluster labels. 
As we can see, not all clusters are 
connected components: some clusters are linked together via the  type 1
outliers. The  type 2 outliers are hidden within clusters.

\paragraph{Type 1 Outlier Detection via $k$-NN}

Type 1 outliers need not be 
isolated 
points in the similarity graph $\mathcal{G}$.
While one may consider handling them by removing vertices of small degree or 
constructing 
the graph by $k$-nearest neighbours\cite{vonLuxburg2007},  it turns out 
that the connectivity of these vertices is much less predictable than that 
required by hard or soft 
threshold. 
A more effective approach to handling Type 1 outliers is after the spectral 
clustering step.

Reexamining Figures \ref{fig:mrt:clusters-hard} and  
\ref{fig:mrt:graph}, note that the Type 1 outliers are such that 
 the 
journey's cluster label is 
different from the lab of its $k$ nearest neighbours at that station. Thus the 
method developed for Type 1 outlier removal is through a $k$-NN approach; see  
Algorithm \ref{alg:mrt:outlier} in the Appendix.
The $k$-NN is not only robust to mis-classified 
journeys, but also  to a having too many clusters $k$. 

\paragraph{Type 2 Outlier Detection via a Mean-Absolute-Deviation Distance 
Metric}

Type 2 outliers cannot be readily  identified by $k$-NN. Hence, a   more 
traditional method was developed to detect these outliers.

\begin{definition}[Median absolute deviation]
    Let $x_1, x_2, \ldots, x_n \in \mathbb{R}$, the \emph{median absolute 
deviation} is defined as
    \[
    \emph{MAD} = \emph{median}_{i = 1}^n{(x_i - \emph{median}_{j = 1}^n{x_j})}.
    \]
\end{definition}
\begin{definition}[Consistent estimator of standard deviation]
    \label{def:mrt:consistent}
    Let $x_1, x_2, \ldots, x_n \sim \mathcal{N}(\mu, \sigma)$ i.i.d\ and 
    \emph{MAD} the median absolute deviation, then 
    \[
    \hat{\sigma} = \frac{1}{\Phi^{-1}(3/4)}\emph{MAD} \approx 1.4826\;\emph{MAD}
    \]
    is a consistent estimator of 
    $\sigma$\cite{doi:10.1080/01621459.1993.10476408}.
\end{definition}
The advantage of using Definition \ref{def:mrt:consistent} as the estimator of 
standard deviation is that it is robust to outliers as long as they are a 
minority. 
 Algorithm \ref{alg:mrt:outlier2} in the Appendix was developed using these two definitions for  Type 2 outlier 
detection.

The clusters that result from the spectral clustering step after outlier removal 
represent the space-time trajectories of each train.  For each  cluster and 
every   station, 
the arrival and departure times are given by  the minimum and maximum timestamps 
in the cluster. 
To estimate trains when intermediate stations have no WiFi observations,   the 
envelope (not 
necessarily convex) of each cluster can be used to connect the arrival and departure 
times at stations with data.

\section{Quantitative Evaluation}
\label{sec:mrt:evaluation}

\begin{figure}[htbp!]
    \centering
    \begin{subfigure}[t]{0.45\textwidth}
        \includegraphics[width = 
\textwidth]{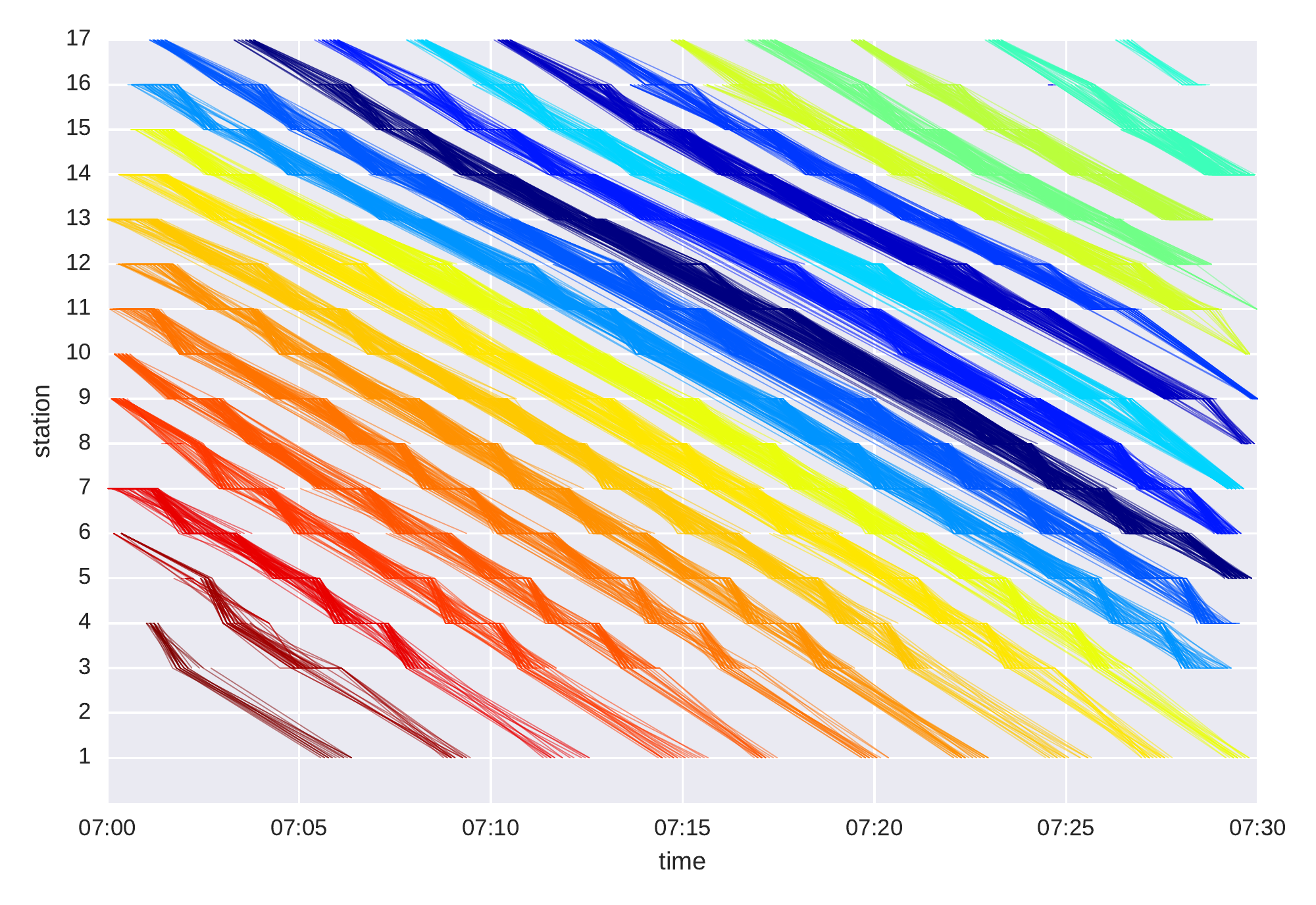}
        \caption{Clusters with outliers removed by baseline}
        \label{fig:mrt:stationwise-denoised}
    \end{subfigure}
    \begin{subfigure}[t]{0.45\textwidth}
        \includegraphics[width = 
\textwidth]{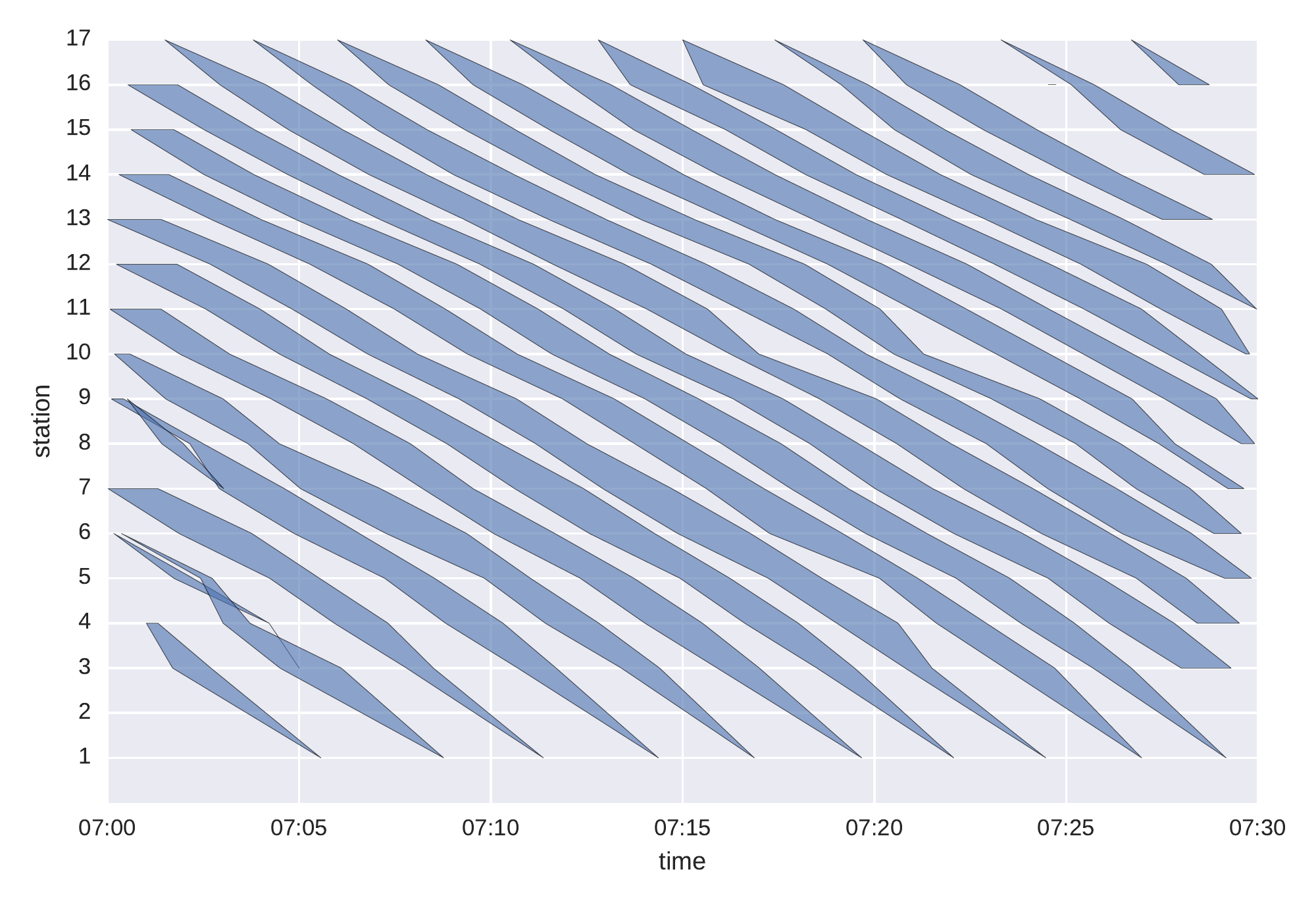}
        \caption{Envelope of trains' arrival and departure point by baseline}
        \label{fig:mrt:stationwise-envelope}
    \end{subfigure}
    \begin{subfigure}[t]{0.45\textwidth}
        \includegraphics[width = 
\textwidth]{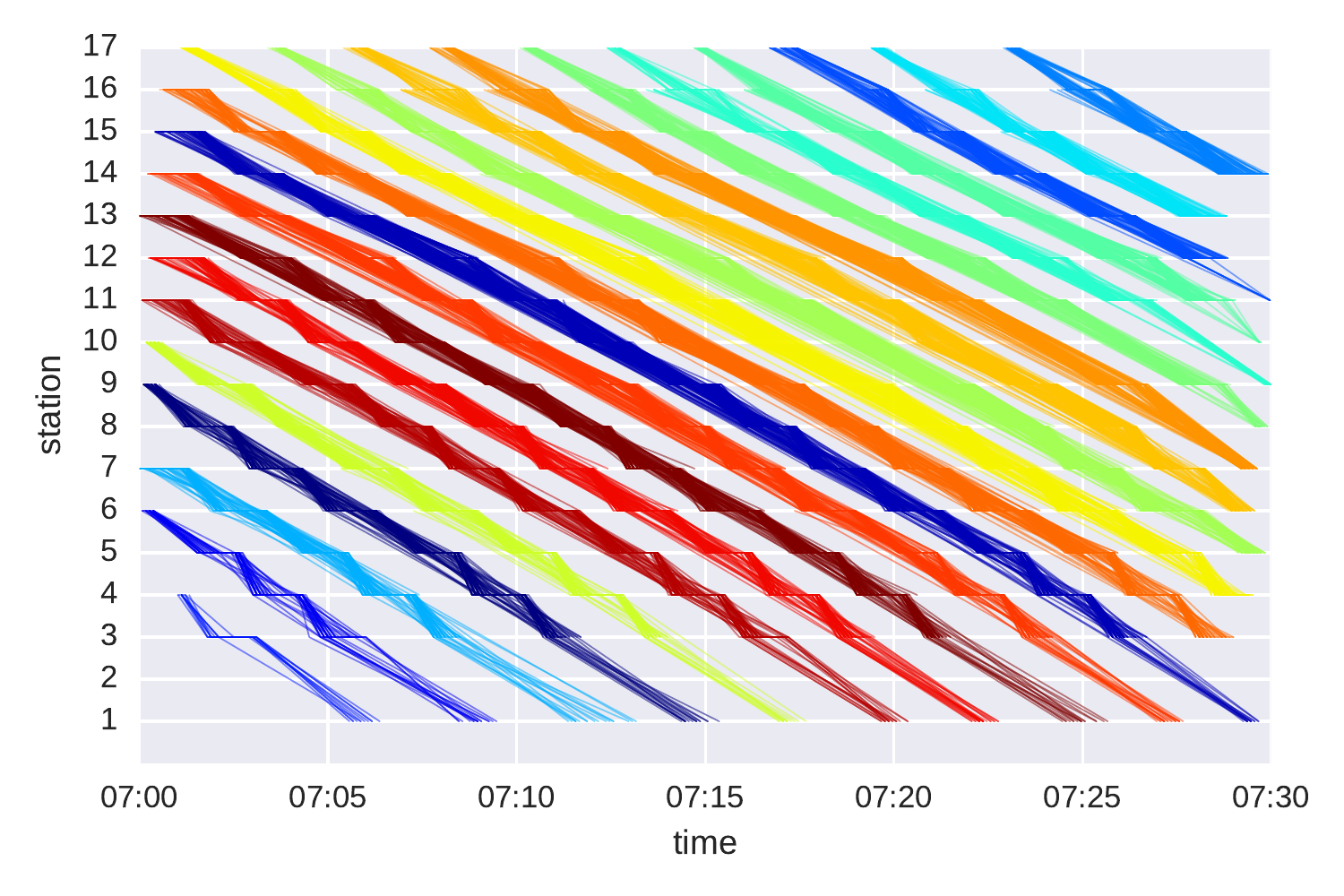}
        \caption{Clusters with outliers removed by spectral clustering}
        \label{fig:mrt:denoised}
    \end{subfigure}
    \begin{subfigure}[t]{0.45\textwidth}
        \includegraphics[width = \textwidth]{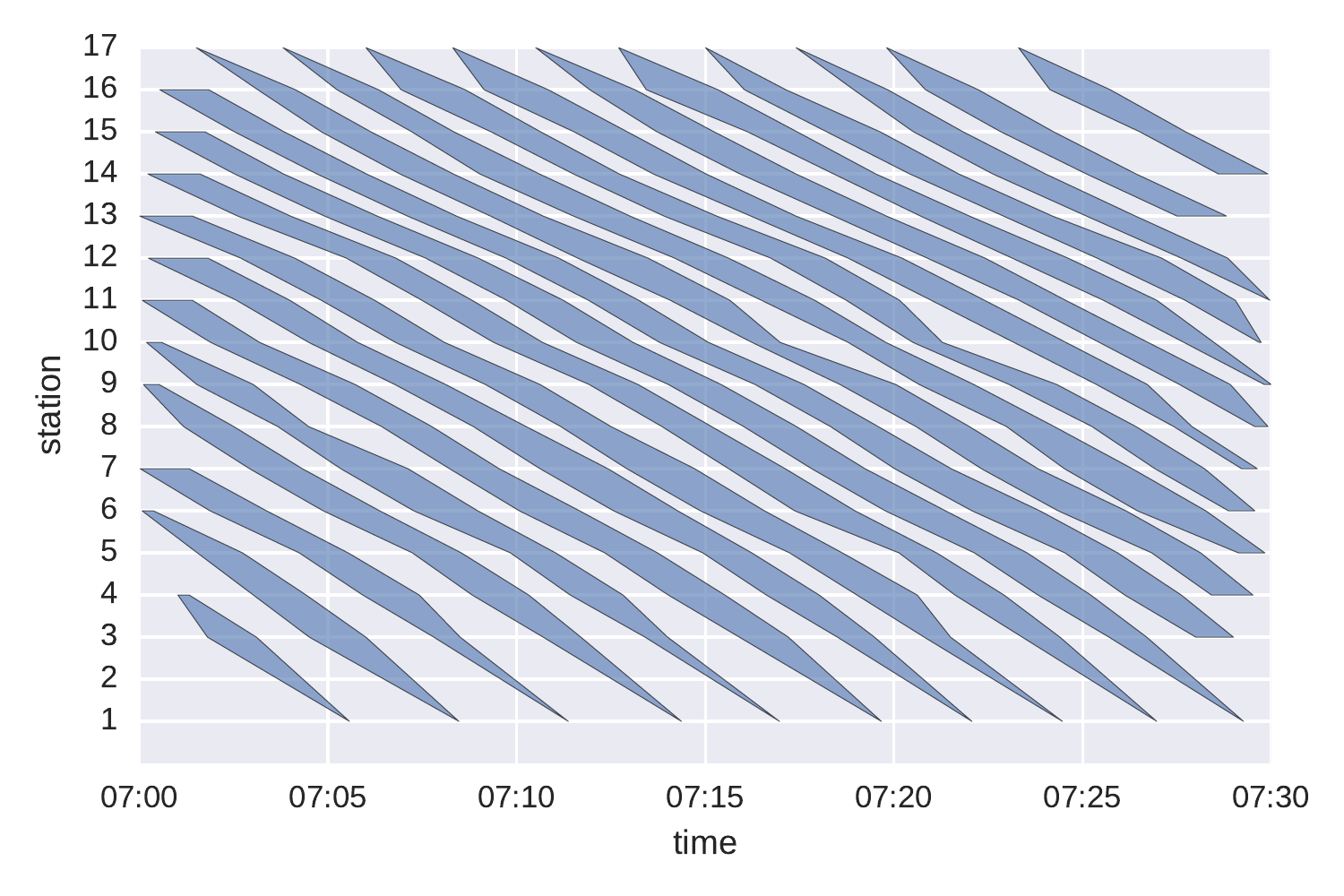}
        \caption{Envelope of trains' arrival and departure point by spectral 
clustering}
        \label{fig:mrt:envelope}
    \end{subfigure}
    \caption{Clusters and trains' arrival and departure by the baseline method 
        and the spectral clustering. Outlier handling is performed using Algorithms 
        \ref{alg:mrt:outlier} and \ref{alg:mrt:outlier2}.} 
    \label{fig:mrt:denoised-envelope}
\end{figure}

The Figures \ref{fig:mrt:denoised-envelope} illustrate a comparison of the baseline method versus the proposed spectral clustering approach. 
The figures show the resulting clusters and  the estimated train trajectories from the envelope using the two methods.
 Despite outlier detection, the baseline method fails to merge two clusters to 
their corresponding trains. 
Neglecting the boundary 
effect\footnote{In practice, it suffice to partition the data into 
overlapped time slots to handle this boundary effect.}
at lower-left and upper-right corner of figure 
\ref{fig:mrt:envelope} into account, the spectral clustering gives   the 
correct number of trains.

To quantitatively evaluate the performance of the methods, we 
compare the result using both the baseline method and the spectral clustering approach with
manual counts of the train arrival times obtained over a period of several hours during peak travel times at four stations over the five workdays during one week.
The results are evaluated quantitatively using the following metrics:
\begin{itemize}
    \item number of trains;
    \item number of  hits, \emph{i.e.}\ number of trains whose 
    error to ground survey is within $\pm$ 1 minute;
    \item hit rate, \emph{i.e.}\ number of hits / number of trains;
    \item Root mean squared error (RMSE) of train arrival time (in minutes).
\end{itemize}
Note that the last three quantities can only be calculated when the number of 
trains is the same as that of the manual counts.

\begin{table}[htbp!]
    \centering
    \begin{tabular}{ccccccc}
        \hline
        Venue&\multicolumn{2}{c}{Manual counts}&\multicolumn{2}{c}{Baseline}&\multicolumn{2}{c}{Spectral 
            Clustering}\\
        \cline{2-7}
        &mean&std&mean&std&mean&std\\
        \hline
        6&40.67&0.58&12.00&1.73&40.33&0.58\\
        8&12.20&5.72&42.60&5.73&42.20&5.72\\
        10&43.75&3.77&44.50&2.89&44.00&3.46\\
        12&42.70&5.17&44.40&5.62&42.40&5.50\\
        13&47.00&0.00&47.33&0.58&47.00&0.00\\
        15&46.00&0.00&45.50&0.71&45.50&0.71\\
        \hline
    \end{tabular}
    \caption{Comparison of number of trains of the baseline method and 
        spectral clustering.}
    \label{tab:mrt:comp:nb_train}
\end{table}

\begin{table}[htbp!]
    \centering
    \begin{tabular}{ccccc}
        \hline
        Venue&\multicolumn{2}{c}{Baseline}&\multicolumn{2}{c}{Spectral 
            Clustering}\\
        \cline{2-5}
        &mean&std&mean&std\\
        \hline
        6 &0.63&NaN&0.89&0.09\\
        8 &0.68&0.54&0.85&0.21\\
        10&0.59&0.53&0.72&0.44\\
        12&0.77&0.23&0.88&0.19\\
        13&1.00&0.00&1.00&0.00\\
        15&1.00&NaN&1.00&0.00\\
        \hline
    \end{tabular}
    \caption{Comparison of hit rate of the baseline method and 
        spectral clustering.}
    \label{tab:mrt:comp:hit_rate}
\end{table}

\begin{table}[htbp!]
    \centering
    \begin{tabular}{ccccc}
        \hline
        Venue&\multicolumn{2}{c}{Baseline}&\multicolumn{2}{c}{Spectral 
            Clustering}\\
        \cline{2-5}
        &mean&std&mean&std\\
        \hline
        6 &1.23&NaN&0.51&0.40\\
        8 &2.55&4.29&0.88&1.55\\
        10&0.71&0.72&0.53&0.60\\
        12&0.46&0.34&0.26&0.25\\
        13&0.00&0.00&0.00&0.00\\
        15&0.00&NaN&0.00&NaN\\
        \hline
    \end{tabular}
    \caption{Comparison of RMSE of the baseline method and 
        spectral clustering.}
    \label{tab:mrt:comp:rmse}
\end{table}

\begin{comment}
\begin{figure}[htbp!]
    \centering
    \begin{subfigure}[t]{.3\textwidth}
        \includegraphics[width = \textwidth]{imgs/comp-nb-trains.pdf}
        \caption{Number of trains}
        \label{fig:mrt:nb_trains}
    \end{subfigure}
    \begin{subfigure}[t]{.3\textwidth}
        \includegraphics[width = \textwidth]{imgs/hit_rate.pdf}
        \caption{Hit rate}
        \label{fig:mrt:hit_rate}
    \end{subfigure}
    \begin{subfigure}[t]{.3\textwidth}
        \includegraphics[width = \textwidth]{imgs/rmse.pdf}
        \caption{RMSE}
        \label{fig:mrt:rmse}
    \end{subfigure}
    \caption{Comparison of hit rate and RMSE of the baseline method and 
        spectral clustering.}
    \label{fig:mrt:comp}
\end{figure}
\end{comment}

Tables \ref{tab:mrt:comp:nb_train}, \ref{tab:mrt:comp:hit_rate} and 
\ref{tab:mrt:comp:rmse} show
the mean number of trains observed manually, and the mean number of trains 
estimated by the baseline method and our proposed spectral clustering approach.

One flaw of the baseline method 
is that it is prone to ``inventing''  trains by creating clusters that do not correspond to any physical train. The spectral clustering approach, as it considers journeys, does not suffer from this problem.
Tables \ref{tab:mrt:comp:hit_rate} and \ref{tab:mrt:comp:rmse} 
show a comparison of the hit 
rate and RMSE of 
both methods. The spectral clustering approach is empirically more stable and 
outperforms the baseline method 
in terms of both hit rate and RMSE. 

\begin{figure}[htbp!]
    \centering
    \includegraphics[width = 
    0.7\textwidth]{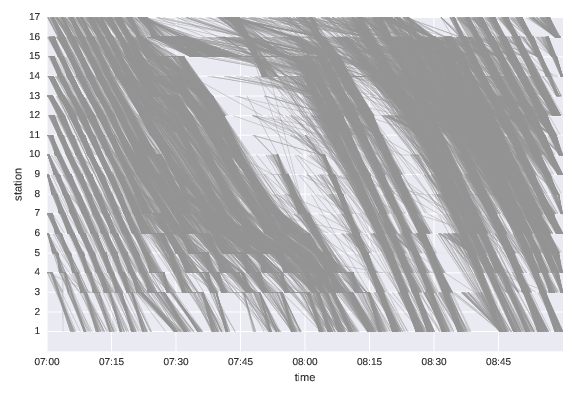}
    \caption{Derived journeys during an incident.}
    \label{fig:mrt:incident}
\end{figure}

To assess the performance of the models during a train disruption, we run the models on the WiFi data obtained during   
a two-hour train incident. Specifically,  a train disruption occurred at station 
$15$ between  7 and 8 am wherein the traffic was  interrupted for half an hour and 
then partially resumed but remained perturbed until $9$am. Shortly after the 
incident, shuttle buses were put into service to help commuters to ``skip'' the affected
station $15$. Figure \ref{fig:mrt:incident} shows the derived journeys. Table \ref{tab:mrt:comp-incident}
shows a comparison of the baseline method and the
spectral clustering approach. The spectral clustering approach again
outperforms the baseline both in terms of hit rate and RMSE.

\begin{comment}
\begin{figure}[htbp!]
    \centering
    \begin{subfigure}[t]{.3\textwidth}
        \includegraphics[width 
        = \textwidth]{imgs/2016-07-29-comp-nb-trains.pdf}
        \caption{Number of trains}
        \label{fig:mrt:nb_trains-incident}
    \end{subfigure}
    \begin{subfigure}[t]{.3\textwidth}
\includegraphics[width=\textwidth]{imgs/2016-07-29-comp-hit-rate.pdf}
        \caption{Hit rate}
        \label{fig:mrt:hit_rate-incident}
    \end{subfigure}
    \begin{subfigure}[t]{.3\textwidth}
        \includegraphics[width=\textwidth]{imgs/2016-07-29-comp-rmse.pdf}
        \caption{RMSE}
        \label{fig:mrt:rmse-incident}
    \end{subfigure}
    \caption{Comparison of number of trains, hit rate and RMSE of the baseline 
method and spectral clustering to ground survey during an incident period.}
    \label{fig:mrt:comp-incident}
  \end{figure}
\end{comment}
  
  \begin{table}[htbp!]
   \centering
   \begin{tabular}{cccccccc}
        \hline
        Venue&\multicolumn{3}{c}{Number of Trains}&\multicolumn{2}{c}{Hit 
rate}&\multicolumn{2}{c}{RMSE}\\
\cline{2-8}
        &Manual counts&Baseline&Spectral&Baseline&Spectral&Baseline&Spectral\\
        \hline
        8&33&33&33&0.06&0.52&7.50&3.63\\
        12&30&30&29&0.66&-&0.84&-\\
        \hline
   \end{tabular}
    \caption{Comparison of number of trains, hit rate and RMSE of the baseline 
method and spectral clustering to ground survey during an incident period.}
    \label{tab:mrt:comp-incident}
  \end{table}

\begin{comment}
\begin{figure}[htbp!]
    \centering
    \begin{subfigure}[t]{0.45\textwidth}
\includegraphics[width=\textwidth]
        {imgs/2016-07-29-station-clusters-denoised.png}
        \caption{Clusters by baseline method}
    \end{subfigure}
     \begin{subfigure}[t]{0.45\textwidth}
        \includegraphics[width=\textwidth]
        {imgs/2016-07-29-station-arrival-departure.pdf}
        \caption{Envelope of trains’ arrival and departure point by baseline 
method}
    \end{subfigure}
    \begin{subfigure}[t]{0.45\textwidth}
        \includegraphics[width=\textwidth]
        {imgs/2016-07-29-clusters-denoised.png}
        \caption{Clusters by spectral clustering}
    \end{subfigure}
     \begin{subfigure}[t]{0.45\textwidth}
        \includegraphics[width=\textwidth]
        {imgs/2016-07-29-arrival-departure.pdf}
        \caption{Envelope of trains’ arrival and departure point by spectral 
clustering}
    \end{subfigure}
    \caption{Incident}
\end{figure}
\end{comment}

\begin{figure}[htbp!]
    \centering
    \includegraphics[width = 0.8\textwidth]{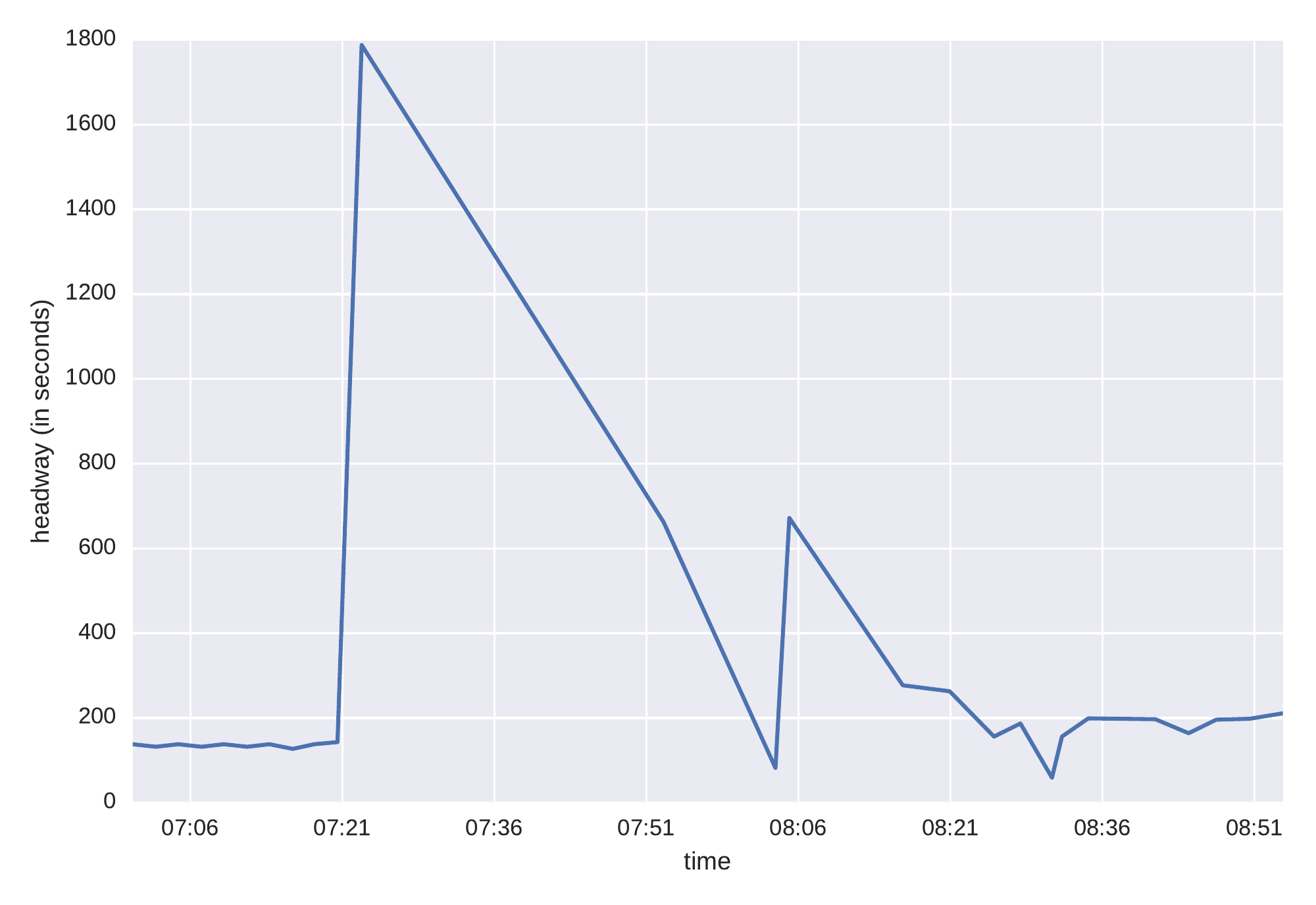}
    \caption{Headways at station $15$ derived from the estimated timetable 
obtained through the spectral clustering approach 
during an incident period.}
\label{fig:mrt:headway}
\end{figure}

Figure \ref{fig:mrt:headway} shows how such a system can be used for very low-latency incident detection in real-time in the public transport network. 
The figure shows a time series of the derived  train headways at station $15$ 
produced by the
spectral clustering approach during the incident period. 
The headways are derived from the estimated timetable as simply taking the 
difference between successive arrivals of the trains into the station. The sharp 
increase in headway is evident in 
the peaks 
shortly after the incident occurs. 

\section{Conclusions}
We have presented a passive WiFi based approach to monitor public transport service levels in real-time. Specifically, we propose a spectral clustering method based on derived commuter trajectories to determine the real-time train timetable, and thus the actual train headways and in-station dwell times. 
 In most cases, there are no other publicly-available sources for this information. Yet, it is indispensable for the real-time monitoring of public transport service levels such as train delays. We showed how the proposed system   can be used for very low-latency incident detection in the public transport system by detecting anomalous headways or in-station dwell times.
 The methods proposed make use of the advantages of the high-frequency WiFi data while minimizing the impact of the noise in the data. 
 
 We considered each line of the train network separately and derived the values of interest by train line. It would be of interest to extend the methodology to handle journeys across multiple lines, i.e. with transfers, as well as multi-modal journeys, if WiFi traces can be retrieved, e.g. from bus stops as well as the train stations. 
A natural extension of this work is also for  the monitoring of crowd-related public transport service levels and disruptions. However, the noise inherent in the passive WiFi data are more challenging as regards the estimation of crowd-related quantities.

\section*{Acknowledgments} The authors gratefully acknowledge the contribution of the Singapore Land Transport Authority (LTA) to the work described in this paper.

\bibliography{reference}{}
\bibliographystyle{unsrt}

\begin{appendices}
\section{Appendix}
\label{sec:appendix}

\begin{algorithm}
    \DontPrintSemicolon
    \SetKwInOut{Input}{input}
    \SetKwInOut{Output}{output}
    \Input{Set of $N$ records $\mathcal{R} = \{(m_i, v_i, t_i)\}_{i = 1}^N$ 
        with $m_i$ MAC address, $v_i$ venue id and $t_i$ timestamp}
    \For{every venue id v in $\mathcal{R}$}{
        set of clusters $\mathcal{C}_v = \{C_{v, -1}, C_{v, 0}, 
        C_{v,1}, \ldots, C_{v, 
            n}\} \longleftarrow$ DBSCAN\;
        remove $C_{v, -1}$\tcp*{DBSCAN gives outliers the label $-1$}
        \For{every cluster $C_{v, i}$}{
            old\_labels = $\varnothing$\;
            \For{every record $r_{v, i, k}$ in $C_{v, i}$}{
                \If{exists $r_{v', i', k'} = r_{v, i, k}$ s.t.\ the difference 
                    of timestamps falls in $[\tau_1, \tau_2]$}{
                    $i_{\mathrm{old}} = i'$\;
                }
                \Else{
                    $i_{\mathrm{old}} = -1$\;
                }
                append $i_{\mathrm{old}}$ to old\_labels\;
            }
            take old\_label as the majority of old\_labels\;
            \tcp{the cluster correspond to a train seen previously}
            \If{old\_label $\neq -1$}{
                change the label of $C_{v, i}$ to old\_label\;
            }
            \tcp{the cluster stands for a new train}
            \Else{
                change the label of $C_{v, i}$ to $\max{\{i| C_{v, i}\}} + 1$\;
            }
        }
    }
    \caption{Baseline method}
    \label{alg:mrt:stationwise}
\end{algorithm}

\begin{algorithm}
    \DontPrintSemicolon
    \SetKwInOut{Input}{input}
    \SetKwInOut{Output}{output}
    \Input{Set of $N$ points $V = \{\bm{t}_i\}_{i=1}^N$\\
        Number of clusters $k$ \\
        $\tau$ for $\textrm{sim}_\textrm{hard}$ or $\sigma$ for 
        $\textrm{sim}_{\textrm{soft}}$}
    %Delete points with $l^0$ norm $1$ (for simplicity, 
    %we still note the number 
    %of points $N$)\;
    Construct graph $\mathcal{G}$ with weight matrix $W$ as in 
    definition \ref{def:mrt:graph}\;
    Remove all isolated points\;
    Compute the unnormalized Laplacian $L$\;
    Compute the eigenvalues and eigenvalues of generalized eigenproblem $Lu = 
    \lambda Du$\;
    Keep only the first $k$ eigenvectors $u_1, \ldots, u_k$\;
    Let $U\in\mathbb{R}^{N\times k}$ be the matrix containing $u_1, \ldots, 
    u_k$ as columns\;
    For $i = 1, \ldots, n$, let $y_i\in\mathbb{R}^k$ be the vector 
    corresponding to the $i$-th row of $U$\;
    Cluster the points $(y_i)_{i = 1, \ldots, n}$ in $\mathbb{R}^k$ with the 
    $k$-means algorithm into $k$ clusters\;
    \Output{$N$ labels $\{l_i\}_{i=1}^N\subset\llbracket 1,k\rrbracket^N$}
    \caption{Normalised spectral clustering\cite{vonLuxburg2007}}
    \label{alg:mrt:spectral}
\end{algorithm}

\begin{algorithm}
    \DontPrintSemicolon
    \SetKwInOut{Input}{input}
    \SetKwInOut{Output}{output}
    \Input{Set of $N$ points $V = \{\bm{t}_i\}_{i=1}^N$\\
        $\tau$ for $\textrm{sim}_\textrm{hard}$ or $\sigma$ for 
        $\textrm{sim}_{\textrm{soft}}$}
    Construct graph $\mathcal{G}$ with weight matrix $W$ as in 
    definition \ref{def:mrt:graph}\;
    Remove all isolated points\;
    Compute the unnormalized Laplacian $L$\;
    Compute the eigenvalues and eigenvalues of generalized eigenproblem $Lu = 
    \lambda Du$\;
    Choose the number of clusters $k$ by eigengap heuristic\\
    Keep only the first $k$ eigenvectors $u_1, \ldots, u_k$\;
    Let $U\in\mathbb{R}^{N\times k}$ be the matrix containing $u_1, \ldots, 
    u_k$ as columns\;
    For $i = 1, \ldots, n$, let $y_i\in\mathbb{R}^k$ be the vector 
    corresponding to the $i$-th row of $U$\;
    Cluster the points $(y_i)_{i = 1, \ldots, n}$ in $\mathbb{R}^k$ with the 
    $k$-means algorithm into $k$ clusters\;
    \Output{$N$ labels $\{l_i\}_{i=1}^N\subset\llbracket 1,k\rrbracket^N$}
    \caption{Normalised spectral clustering with adaptive number of clusters}
    \label{alg:mrt:spectral2}
\end{algorithm}

\begin{algorithm}
    \DontPrintSemicolon
    \SetKwInOut{Input}{input}
    \SetKwInOut{Output}{output}
    \Input{$N$ records $\mathcal{R} = \{r_i\}_{i=1}^N = \{(m_i, v_i, t_i, 
        c_i)\}_{i=1}^N$ with $m_i$ MAC address, $v_i$ venue id, $t_i$ timestamp 
        and 
        $c_i$ cluster label;\\
        number of neighbours $k$
    }
    \For{every venue id $v$ in $\mathcal{R}$}{
        $\mathcal{R}_v \longleftarrow \{r_i\in\mathcal{R}|v_i = v\}$\;
        use $k$-NN to assign a new cluster label $c_v^{\textrm{neightbour}}$ 
        to every record $r_v$\;
    }
    $\mathcal{R}'\longleftarrow \{r_i \in \mathcal{R}| c_i = 
    c_i^{\textrm{neightbour}}\}$\;
    \Output{$\mathcal{R}'$}
    \caption{ Type 1 outlier detection using  $k$-NN}
    \label{alg:mrt:outlier}
\end{algorithm}

\begin{algorithm}
    \DontPrintSemicolon
    \SetKwInOut{Input}{input}
    \SetKwInOut{Output}{output}
    \Input{$N$ records $\mathcal{R} = \{r_i\}_{i=1}^N = \{(m_i, v_i, t_i, 
        c_i)\}_{i=1}^N$ with $m_i$ MAC address, $v_i$ venue id, $t_i$ timestamp 
        and 
        $c_i$ cluster label;\\
        threshold $\tau$
    }
    \For{every venue id $v$ in $\mathcal{R}$}{
        $\mathcal{R}_v \longleftarrow \{r_i\in\mathcal{R}|v_i = v\}$\;
        \For{every label $c$}{
            $\mathcal{R}_{v,c} \longleftarrow \{r_i\in\mathcal{R}_v|c_i = c\}$\;
            \emph{MAD} $\longleftarrow \textrm{MAD}(t_{v,c})$\;
            $\hat{\sigma} = 1.4826\textrm{MAD}$\;
            \For{every label $r_{v,c}$}{
                \If{$|t_{v,c} - \textrm{MAD}| \geq \tau\hat{\sigma}$}{
                    remove $r_{v,c}$\;
                }
            }
        }
    }
    \Output{$\mathcal{R}'$}
    \caption{ Type 2 outlier detection using  Median Absolute Deviation}
    \label{alg:mrt:outlier2}
\end{algorithm}
\end{appendices}
\end{document}